\documentclass[aps,prl,twocolumn,groupedaddress,superscriptaddress,showpacs]{revtex4-1}

\usepackage{amsmath}
\usepackage{graphicx}
\usepackage{float}
\usepackage{color}

\begin{document}

\title{Self-assembly of colloidal bands driven by a periodic external field} 

\author{Andr\'e S. Nunes}
\affiliation{Departamento de F\'{\i}sica, Faculdade
de Ci\^{e}ncias, Universidade de Lisboa, P-1749-016 Lisboa, Portugal, and
Centro de F\'isica Te\'orica e Computacional, Universidade de Lisboa,
P-1749-016 Lisboa, Portugal}

\author{Nuno A. M. Ara\'ujo}
\email{nmaraujo@fc.ul.pt}
\affiliation{Departamento de F\'{\i}sica, Faculdade
de Ci\^{e}ncias, Universidade de Lisboa, P-1749-016 Lisboa, Portugal, and
Centro de F\'isica Te\'orica e Computacional, Universidade de Lisboa,
P-1749-016 Lisboa, Portugal}

\author{Margarida M. Telo da Gama}
\affiliation{Departamento de F\'{\i}sica, Faculdade
de Ci\^{e}ncias, Universidade de Lisboa, P-1749-016 Lisboa, Portugal, and
Centro de F\'isica Te\'orica e Computacional, Universidade de Lisboa,
P-1749-016 Lisboa, Portugal}

\begin{abstract}
We study the formation of bands of colloidal particles driven by periodic
external fields. Using Brownian dynamics, we determine the dependence of the
band width on the strength of the particle interactions and on the intensity
and periodicity of the field. We also investigate the switching (field-on)
dynamics and the relaxation times as a function of the system parameters. The
observed scaling relations were analyzed using a simple dynamic
density-functional theory of fluids.
\end{abstract}

\maketitle

\section{Introduction}

The possibility of obtaining materials with enhanced physical properties from
the self-assembly of colloidal particles has motivated experimental and
theoretical studies over
decades~\cite{Lash15,Blaaderen1995,Blaaderen2006,Kim2011}. Initially, the focus
was on identifying novel phases and constructing the equilibrium phase
diagrams, based on the properties of the individual particles (shape, size, and
chemistry)~\cite{Damasceno12,Sacana2013,Dias2013}. However, the impressive
advance of optical and lithographic techniques opened the possibility of
exploring alternative routes as, for example, the use of substrates and
interfaces~\cite{Araujo07,Araujo08,Garbin13,Dias2013b,Cademartiri15,Araujo15},
the control of the suspending medium~\cite{Silvestre2014} or the assembly under
flow~\cite{Klapp2015Soft}. More recently, there has been a sustained interest
on the self-assembly under the presence of an electromagnetic (EM)
field~\cite{Furst2014,Furst2015,Klapp2015PRE,LowenRev,Dobnikar}.  Uniform EM
fields couple to the rotational degrees of freedom of magnetic aspherical
particles allowing to control their orientation~\cite{Davies14} or fine-tune
the strength and directionality of the particle
interactions~\cite{LowenRev,Dobnikar,Keim2013}. Space-varying periodic fields
are employed to impose constraints on the particle position, forming virtual
molds that induce spatial periodic
patterns~\cite{Demirors2013,Egelhaaf,Randall2009}.

Under external constraints, colloidal suspensions are usually driven out of
equilibrium and the relaxation towards equilibrium results from the competition 
of various mechanisms occurring at different length and time scales. Thus, besides
the identification of the equilibrium structures and their dependence on the
experimental conditions, it is of paramount practical interest to characterize
the kinetic pathways towards the desired structures and the timescales
involved. Here, we study the prototypical example of field-driven self-assembly
of colloidal particles. For simplicity, we consider the formation of colloidal
bands driven by a periodic (sinusoidal) electromagnetic field and study how the
equilibrium structures and the dynamics towards equilibrium depend on the field
strength and periodicity, as well as on the particle interactions. We
perform extensive Brownian dynamics simulations and complement the study with a
coarse-grained analysis based on the dynamic density-functional theory of
fluids (DDFT).

The paper is organized in the following way. In Section~\ref{sec::model}, we
present the model and simulation details. The dependence of the stationary state
structure and relaxation dynamics on the model parameters is discussed in
Section~\ref{sec::results}. Finally, we draw some conclusions in
Section~\ref{sec::conclusions}.

\section{Model and Simulations~\label{sec::model}}
We consider a two dimensional system of colloidal particles in the overdamped
regime. The pairwise particle interactions are described by the
repulsive Yukawa potential,
\begin{equation}
V_{ij}(r)=V_{0}\frac{\exp\left(-\alpha r\right)}{r},\label{Yukawa}
\end{equation}
where $r=\vert\vec{r}_i-\vec{r}_j\vert$ is the distance between particles $i$ and
$j$. $V_0$ sets the energy scale and the screening parameter $\alpha$ sets the
range of the potential. A characteristic particle radius, $r_p$, can be defined
as $r_p=(2 \alpha)^{-1}$, which we set as the unit of length. For simplicity, we
consider only repulsive interactions ($V_0>0$). To quantify
the pairwise interactions we define the integrated parameter $A=\int
V_{ij}(\vec{r})d\vec{r}$ expressed in units of $k_BTr^2_p$. In the numerical
simulations we set $\alpha= 1/2$ and change $A$ by varying $V_0$.

We consider an external field that is constant along the $y$-direction and
periodic along the $x$-direction. The particle/field interaction is described by
a periodic potential,
\begin{equation}
V_{ext}(x,y) = V_{ext}(x) = V_E\sin\left(\kappa x\right),\label{external}
\end{equation}
where $V_E$ is the strength of the potential and 
\begin{equation}
\kappa=\frac{2\pi P}{L}\label{omega}
\end{equation}
is the wave number. $P$ sets the number of minima of the potential in a system
of linear length $L$. 

We perform Brownian dynamics simulations where the equation of motion of the
colloidal particle $i$ is
\begin{equation}
  \gamma \frac{d\vec{r}_i}{dt} = -\vec{\nabla} _i\left[\sum_{j}
V_{ij}(r)+V_{ext}(\vec{r}_i)\right]+\vec{\xi}_{i},\qquad j\neq
i\label{MotionEq}
\end{equation}
where $\gamma$ is the Stokes friction coefficient and the last term on the
right-hand side $\vec{\xi}_{i}(t)$ is the Langevin force  mimicking the
particle/fluid interaction, which is randomly sampled from a Gaussian
distribution of zero mean and second moment $\left\langle
\xi_i^k(t)\xi_i^l(t')\right\rangle =2k_BT\gamma\delta_{kl}\delta(t-t')$. The
indices $k$ and $l$ run over the spatial dimensions, $k_B$ is the Boltzmann
constant and $T$ the thermostat temperature of the suspending fluid. Thus, the
random force is uncorrelated in time and space.

We start our simulations with a uniform (random) distribution of colloidal
particles and switch on the field at $t=0$. The simulations were performed
inside square boxes of three different linear lengths $L=\{100,120,150\}$, in
units of the particle radius. By fixing the density $\rho=0.0696$, we simulate 
$N=\{696,1000,1560\}$ colloidal particles, respectively. 

Hereafter, the strength of the external potential $V_E$ is expressed in units of
$k_BT$ and time is defined in units of the Brownian time $\tau=r^2_p\gamma
(k_BT)^{-1}$, which is the time over which a colloidal particle diffuses over a
region equivalent to its area. To integrate the stochastic differential
equations of motion we followed the scheme proposed by Bra\'nka and
Heyes\cite{Branka1999}, which consists of a second-order stochastic Runge-Kutta
scheme, with a time-step of $\Delta t=10^{-4}\tau$. 

\section{Results~\label{sec::results}}
\begin{figure}[t]
  \centering
  \includegraphics[width=\columnwidth]{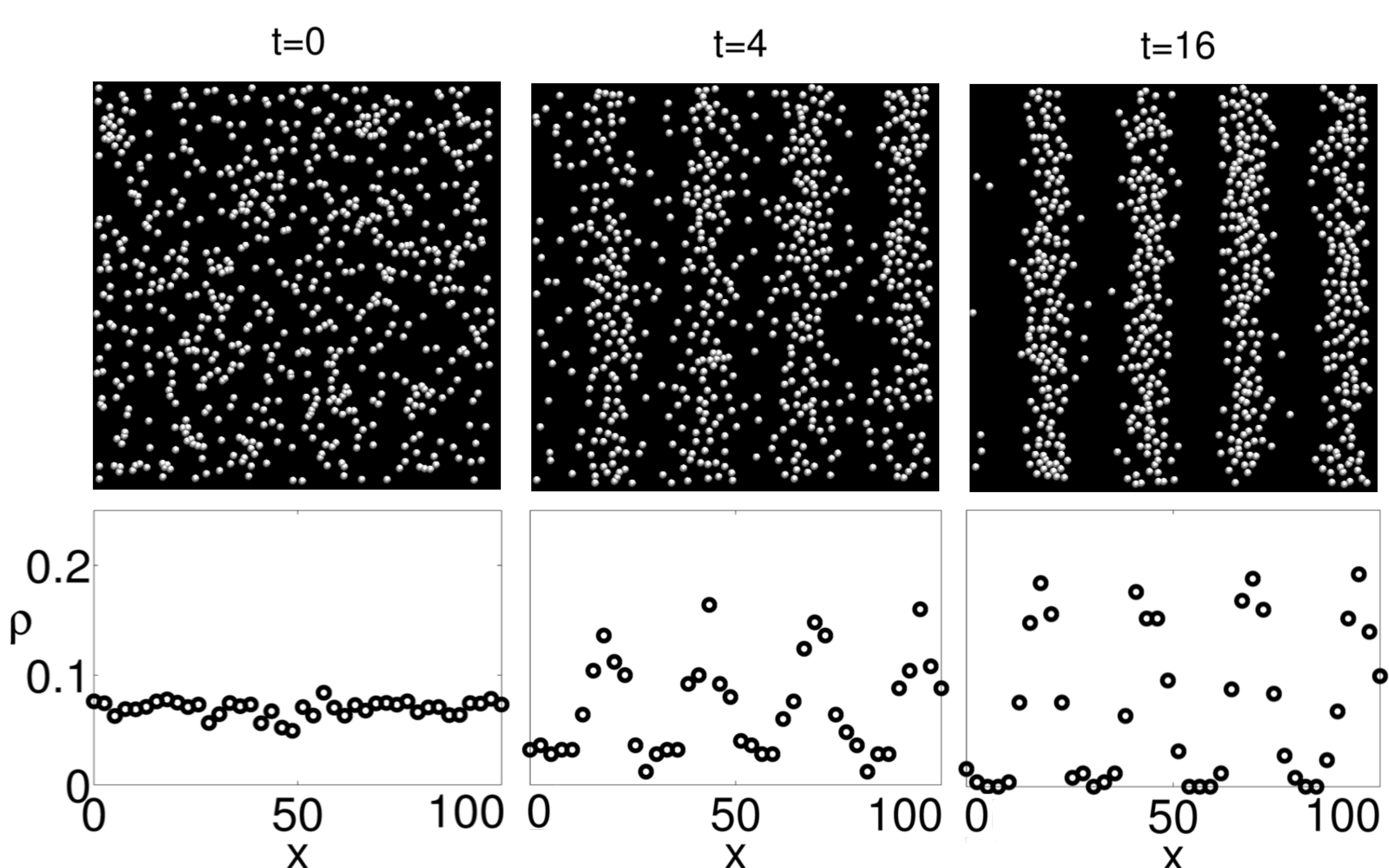}
  \caption{Snapshot of the system for $t=0$, $t=4$ and $t=16$. The plots
correspond to the density profiles along the $x$-direction. Under the influence of
the external potential the particles organize into bands. The linear length of
the system is $L=100$, the interaction parameters are $V_0=5$ and $\alpha=1/2$.
The strength of the field is $V_E=5$ and the number of minima is
$P=4$.}\label{fgr:evol}
\end{figure}

In the absence of external fields the spatial distribution of the colloidal
particles is homogeneous. When the external field is switched on the particles
are dragged towards the closest minimum and self-organize into individual bands
around it (see snapshots in Fig. \ref{fgr:evol}). To quantify the spatial
arrangement of the colloidal particles we measure the density, $\rho(\vec{r})$,
defined as the number of particles per unit area. Given the symmetry of the
external field (see Eq. (\ref{external})), we focus on the profile of the
density along the $x$-direction. Figure \ref{fgr:evol} shows a snapshot and the
density profile for a system with $L=100$ at three different times. The density
profile has the symmetry of the external potential. The maxima of the density
correspond to the minima of the potential. Similarly, the density vanishes at
the maxima of the potential. We run the simulations for $6 \times 10^6$
timesteps. After this time, the changes in the density profile are within the
error bars and thus we consider that the stationary state has been reached.
Simulations with different system sizes  were preformed and we found no
significant finite-size effects.

The collective dynamics and the band structure result from a complex interplay
of the particle/fluid, particle/field and particle/particle interactions. Here,
we analyze how the final structure and dynamics depend on the model parameters.
We start with an analysis of the stationary state and proceed with the study of
the dynamics.

\subsection{Stationary State}
\begin{figure*}[t]
 \centering
 \includegraphics[width=\columnwidth]{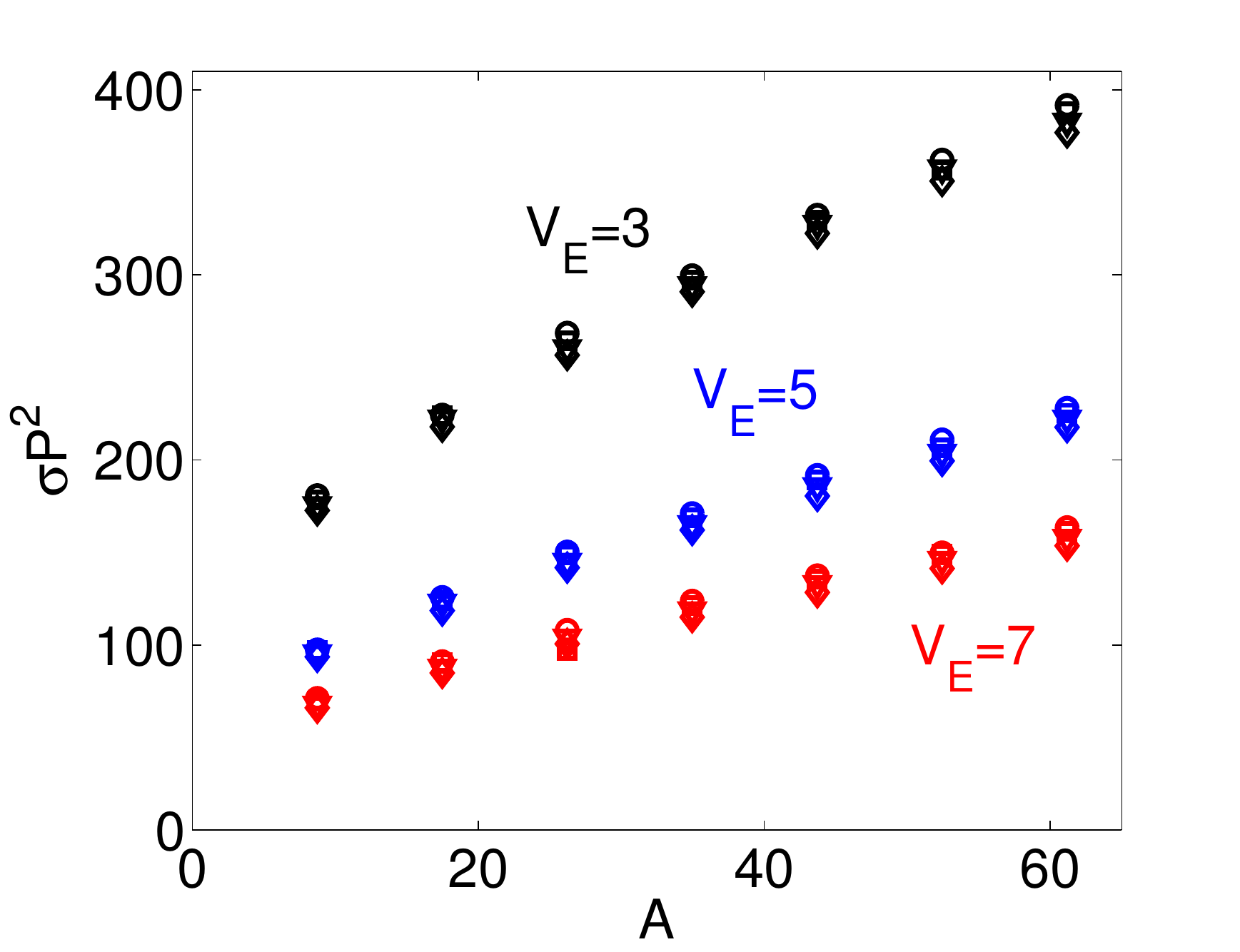} \includegraphics[width=\columnwidth]{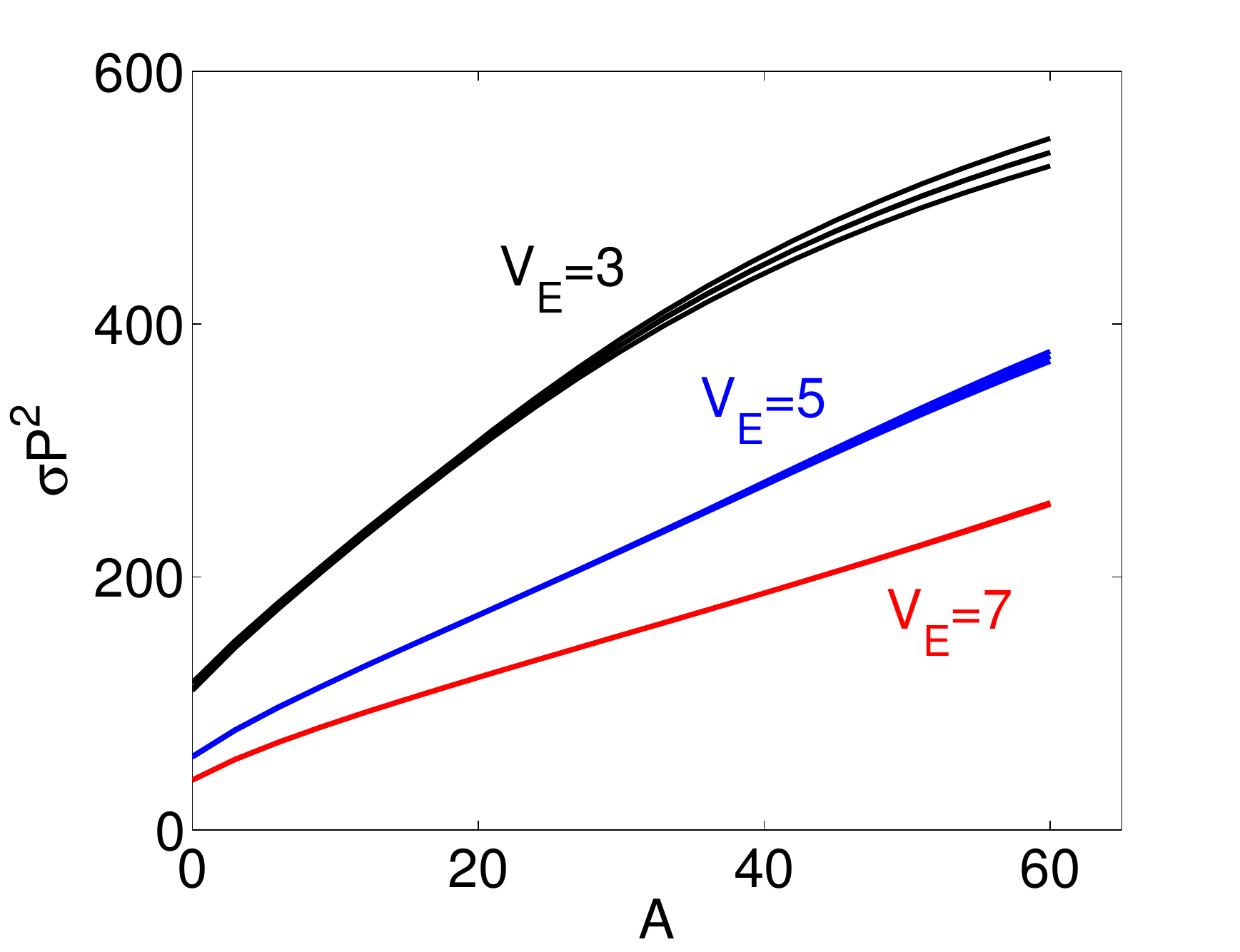}
 \caption{Mean square displacement of the particle position around the potential
minima. On the left-hand side are the results obtained from the Brownian
dynamics simulations and on the right-hand side those from the DFT
calculations. The data is obtained by averaging over $25$ samples rescaled by
$P^{-2}$. Black, blue and red represent respectively $V_E=\{3,5,7\}$ and the
circle, square, triangle and diamond symbols correspond respectively to $P=\{1,2,3,4\}$.
The error bars are smaller than the size of the symbols. The particle/particle
and particle/field interactions have opposite effects on the final
configuration. While the former promotes wider bands the latter favors thinner
ones. \label{fgr:sigmaSim}}
\end{figure*}

\subsubsection{Numerical simulations.}
We first characterize the stationary state that corresponds to the thermodynamic
equilibrium configurations. For $P > L / 2$ the average particle/field
interaction vanishes, as the wave length of the potential is shorter than the
particle diameter. The limit of no external field is then recovered. For $P < L
/ 2$ and strong enough field, particles accumulate around the minima forming a
band structure. Thermal noise and particle/particle interaction cause a
broadening of the bands, giving them an effective thickness. This thickness is a
property of the equilibrium configuration. To quantify it, we measure the mean
square displacement of particles around the local minimum
\begin{equation}
\sigma=\langle x_b^2\rangle - \langle x_b\rangle ^2
\end{equation}
where the average is over particles in the same band, i.e., in between two
consecutive maxima of the potential. $x_b$ is their position relative to the
local minimum along the $x$-direction. The average displacement is zero as the
distribution of particles is symmetrical with respect to the center of the band
(minimum of the potential).

Figure \ref{fgr:sigmaSim} (left panel) shows the dependence of $\sigma$ on the
model parameters. One clearly sees that, for a given thermostat temperature, the
thickness of the bands results from the competition between particle/field and
particle/particle interactions. The particle/field interaction promotes the
formation of thin bands. The stronger the field ($V_E$) the thinner the band is.
By contrast, the particle/particle repulsion tends to homogenize the spatial
distribution of the particles and favors wider bands. Consequently, $\sigma$
increases monotonically with $A$. Note that when we rescale $\sigma$ by $P^2$ a
data collapse is obtained for different $P$ values into a single curve that
depends only on $V_E$ and $A$. In the next section we perform a density-functional 
theory (DFT) analysis to study these dependences.

\subsubsection{Density-functional theory analysis.}
An approximation for the Helmholtz free energy functional can be written as
\begin{equation}
\begin{split}
&\mathcal{F}[\rho]=\int
k_{B}T\rho(\vec{r})\left[\log\left(\rho(\vec{r})\Lambda^{2}\right)-1\right]d\vec{r}
\\ &+\frac{1}{2}\int\int \rho(\vec{r}) \rho(\vec{r}')
V_{pp}(\vec{r}-\vec{r}')d\vec{r}'d\vec{r} \\
&+\int\rho(\vec{r})V_{ext}(\vec{r})d\vec{r}.
\end{split} \label{funtional}
\end{equation}
On the right-hand side, the first term is the free energy of the ideal gas where
$\Lambda$ is the thermal de Broglie wavelength, the second term is the
mean-field approximation to the contribution from the interactions and the last
term is the external potential contribution. We use the local density
approximation (LDA) by setting $\rho(\vec{r}')\simeq\rho(\vec{r})$ which is a
good assumption if the density is a smooth function of the position.

\begin{figure}[ht]
\centering
  \includegraphics[width=\columnwidth]{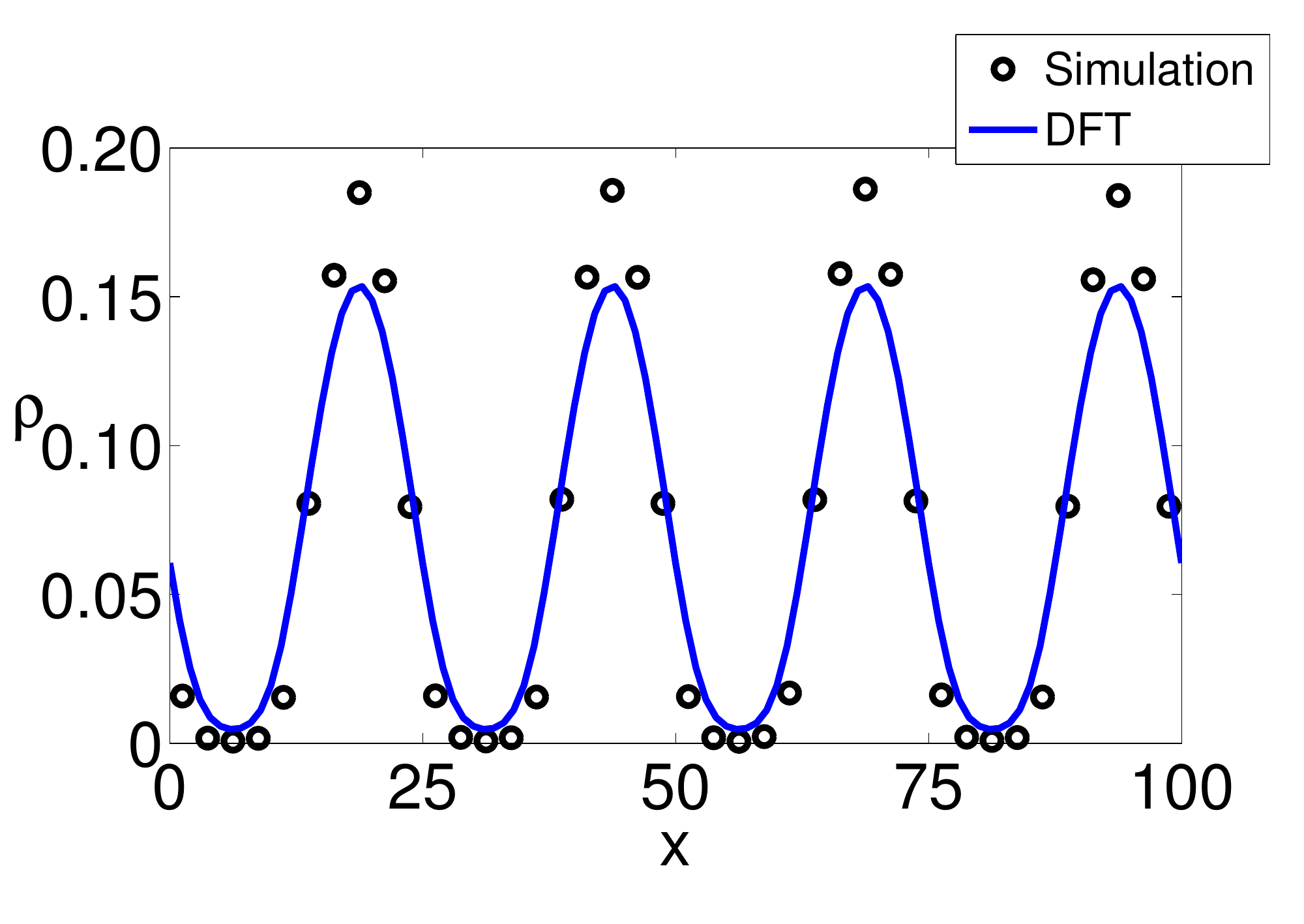}
  \caption{Density profile for $L=100$, $V_E=5$, $A=43.71$ and $P=4$. The dots
are the numerical results obtained by averaging over $100$ samples and the line
is the result of DFT. The theory is in good agreement with the simulations but it
underestimates the density at the maxima. \label{fgr:densityprofile}}
\end{figure}

Now we employ the definition of chemical potential,
$\mu=\frac{\delta\mathcal{F}[\rho]}{\delta\rho(\vec{r})}$, to obtain
\begin{equation}
k_{B}T\log\left(\rho(\vec{r})\Lambda^{2}\right)+A\rho(\vec{r})+V_{ext}(\vec{r})=\mu.
\end{equation}
where $A$ is the interaction parameter defined previously. Rearranging the last equation we find an expression for the
density profile
\begin{equation}
\rho(x)=Z\exp[-\beta\left(V_E \sin(\kappa x)+A\rho(x)\right)] \label{density}
\end{equation}
where $\beta=(k_BT)^{-1}$ and $Z=\Lambda^{-2}e^{\beta\mu}$ is a normalization
constant that represents the density of an ideal gas with chemical potential
$\mu$. Recall that from the symmetry of the external potential the density is
only a function of the $x$-coordinate. To solve this equation we impose an
additional constraint $N=\int\rho(\vec{r})d\vec{r}$ arising from the
conservation of the total number of particles. 

\begin{figure}[ht]
\centering
  \includegraphics[width=\columnwidth]{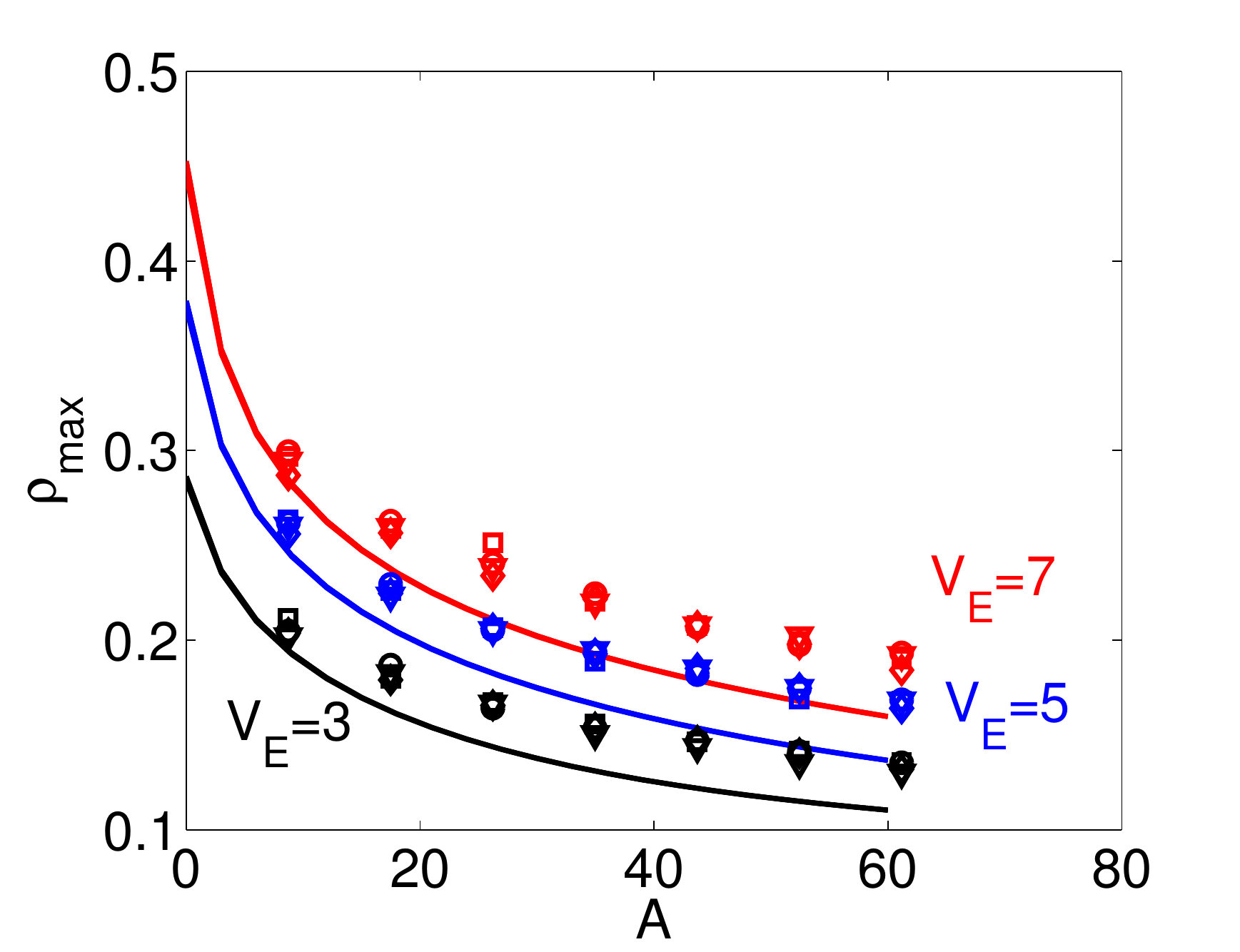}
  \caption{Maximum density of the profile. Each data point is an average
over 25 samples and the lines are the DFT results. Black, blue and red represent
respectively $V_E=\{3,5,7\}$ and the circle, square, triangle and diamond
symbols correspond respectively to $P=\{1,2,3,4\}$. The error bars are smaller than the
size of the symbols. Note that the density does not depend on $P$. The
deviation of the theoretical curves from the simulation results increases
particle interactions increase. The effect of $A$ and $V_E$ on the maximal density
is opposite to their effect on the mean square displacement.
\label{fgr:denmax}}
\end{figure} 
Figure \ref{fgr:densityprofile} shows the density profile obtained from
numerical simulation and DFT for the same set of parameters. The DFT results
deviate from the simulation only at the maximal and minimal densitities. The
maximal density is underestimated and the minimal is slightly
overestimated. We evaluated how this deviation depends on the model parameters.
Figure \ref{fgr:denmax} shows the dependence on $A$ of the maximal density
$\rho_{max}$, defined as the density in the center of the band, for different
values of $V_E$ and $P$. We find that, while the deviation of the DFT
calculation from the simulation increases slightly with $A$ it does not
vary significantly with $V_E$ and $P$. 
This also leads to theoretical values of the mean square displacement $\sigma$, which are larger than those obtained 
by simulation. However, as
shown in Fig. \ref{fgr:sigmaSim} (right panel) one recovers the same qualitative
dependence on the model parameters and the values of $\sigma$ are of the
same order of magnitude. 

Since the density profile is symmetric with respect to the center of the band, the
mean square displacement is given by,
\begin{equation}
\sigma=L\int_{-\frac{L}{2P}}^{\frac{L}{2P}}x^{2}\rho^*(x)dx.
\end{equation}
where $\rho^*(x)=\rho (x)N^{-1}_b$ is the probability density function for
particles in the band and $N_b = L
\int_{-\frac{L}{2P}}^{\frac{L}{2P}}\rho(x)dx$. The $L$ outside of the integral
accounts for the integration of the domain along the $y$-direction. Assuming
that $\rho(x)$ is constant and $N_b=N/P$,
\begin{equation}
\sigma = \frac{L^{2}}{12 P^{2}}\ \ .\label{sigmatheory}
\end{equation}
Thus, $\sigma \sim P^{-2}$ as observed both from the simulations and the DFT
analysis. Note that the maximal density (in Fig. \ref{fgr:denmax})
does not depend on $P$. From Eq. (\ref{density}) we also conclude that, near the
minima, the external potential and the interaction terms have different signs. Consequently, $V_E$ and $A$ have opposite effects on
$\sigma$, as observed in the simulations and DFT calculations, see Fig.
\ref{fgr:sigmaSim}.

\begin{figure}[t]
\centering
  \includegraphics[width=\columnwidth]{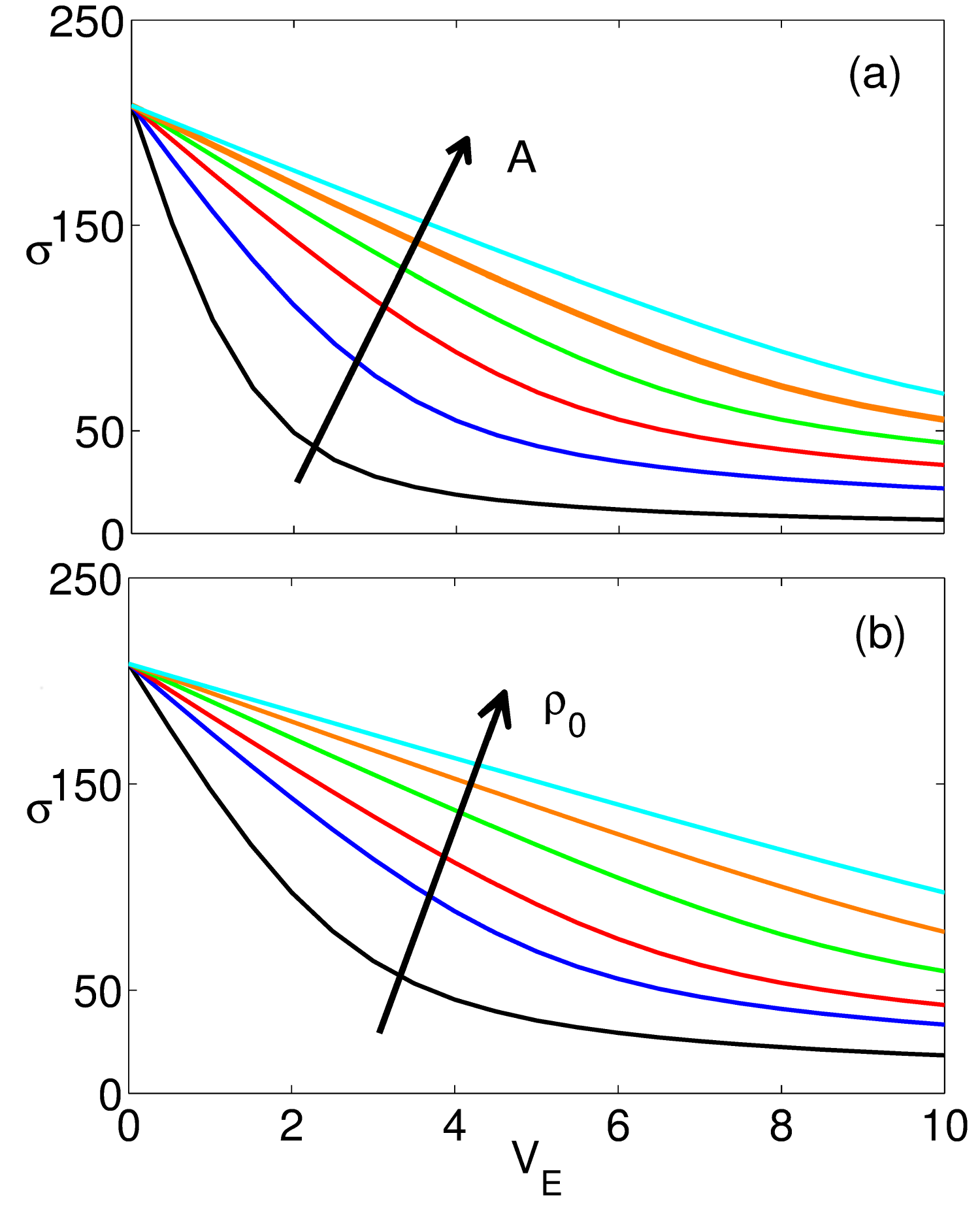}  
  \caption{DFT results for the mean square displacement for $P=2$. The lines in
(a) are for $A=\{0,20,40,60,80,100\}$ and the lines in (b) are for different
initial densities $\rho_0=\{2.50, 6.96, 10.0, 15.0, 20.0, 25.0\}\times10^{-2}$.
For strong inter-particle interactions and weak external potentials the
dependence is linear in $V_E$. The linear dependence is also observed at high
densities. \label{fgr:sigmaE}}
\end{figure}

By taking the logarithm of Eq. (\ref{density}) we obtain
\begin{equation}
\log(\rho(x))+\beta A\rho(x) = log(Z) - \beta V_E sin(\kappa x).
\end{equation}
The dependence of the first term of the left-hand side on the density can be
neglected with respect to the second term in two limiting cases: for strong
particle/particle interactions or high enough densities. In the former, the
strong particle/particle repulsion hinders the formation of bands and
homogenizes the density. The external potential acts as a perturbation to
the uniform distribution of particles, which is sinusoidal in space and linear on
$V_E$. The contributions from the external potential and particle interactions are 
given by $\rho (x)$ which means that the dependence of the mean square displavement 
on the strength of the potential intensity is the same in both limits, 
as seen in Fig. \ref{fgr:sigmaE}(a) and (b). 
For weak particle interactions the curves are linear when the strength of the potential 
is also small. But as the particle/particle interaction increases, 
the potential intensity where the dependence deviates from linear 
also increases. Note that the curve for
$A=0$ corresponds to the ideal gas where the dependence of $\rho(x)$ on the
strength of the potential is exponential as expected from Eq.
(\ref{density}). For $V_E=0$ the density is uniform and the value of $\sigma$ is
given by Eq. (\ref{sigmatheory}) and is the same for all curves.

\subsection{Relaxation Dynamics}
\subsubsection{Numerical simulations.}
We analyze now the relaxation towards the stationary state shown in Fig.
\ref{fgr:evol}. Namely, we investigate how the relaxation time depends on
the model parameters.

\begin{figure}[t]
\centering
  \includegraphics[width=\columnwidth]{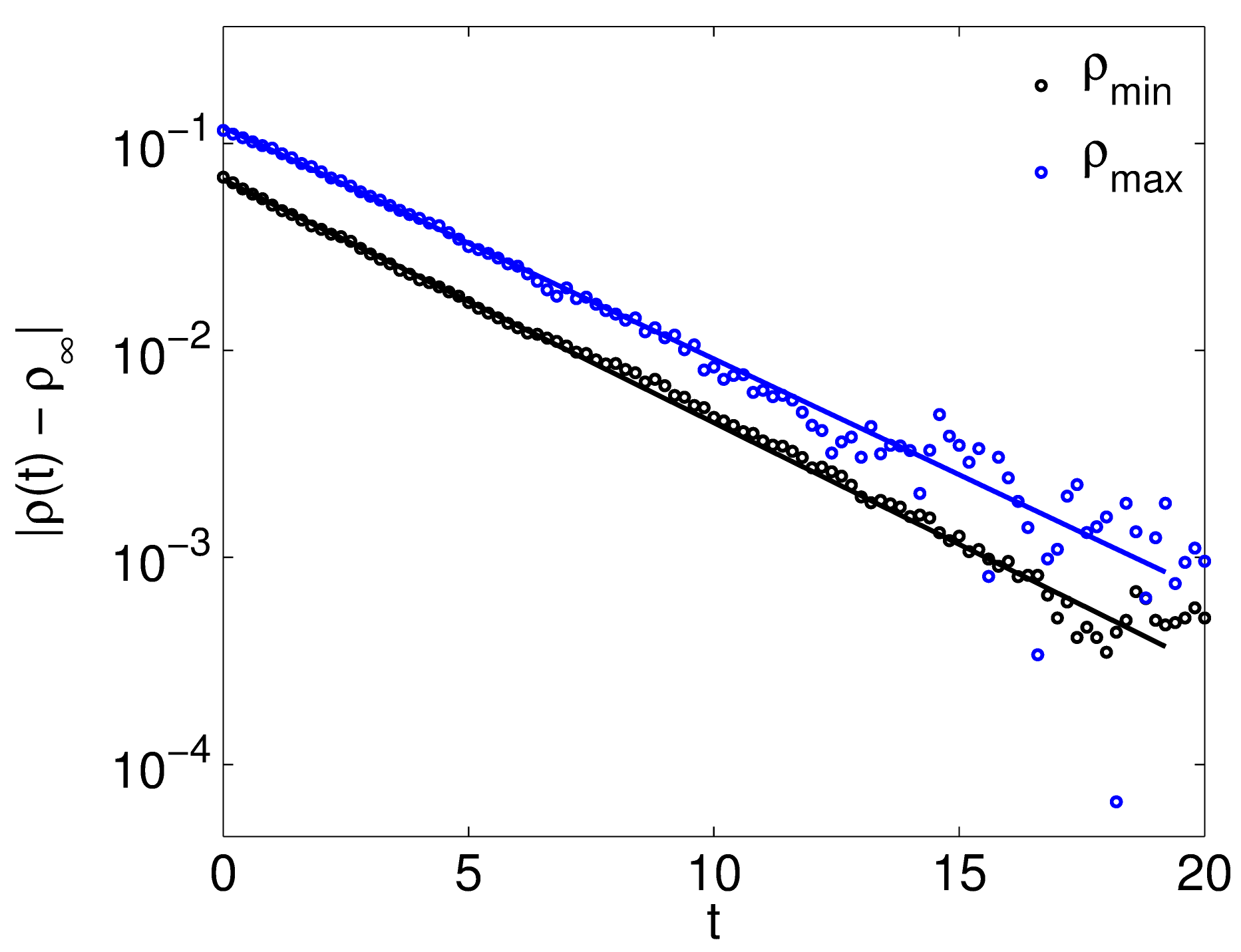}
  \caption{Evolution of the maximal and minimal densities. The relaxation
into the stationary state is exponential. The scattered data are the numerical
results and the lines the exponential fitting. The results shown are averages
over $100$. The parameters are $L=100$, $V_E=5$, $A=43.71$ and $P=4$.
\label{fgr:density_t}}
\end{figure}

We start with a uniform spatial distribution of particles, $\rho(x)=\rho_0$. As
the external potential is switched on the colloidal particles move accordingly
and regions of high density of colloids are formed (bands). For a systematic
analysis, we focus on the evolution of the positions of the minima,
$\rho_{min}$, corresponding to the potential maxima, and the positions of
maxima, $\rho_{max}$, corresponding to the potential minima.
Figure \ref{fgr:density_t} shows the convergence of these densities to their
value in the stationary state, $\rho_{\infty}$, for a particular set of
parameters. One clearly sees an exponential decay in time. 

We define the relaxation time $\tau^{relax}$ as the inverse of the slopes
of the evolution curves. Figures \ref{fgr:taumin} and \ref{fgr:taumax} show that
the dependence on the model parameters is different for the minima and the
maxima suggesting that the underlying relaxation mechanisms are different. In general,
strong potentials favor short relaxation times while the dependence
on the particle/particle interaction is not straightforward. Also, 
data collapse is observed when we rescale $\tau$ with $P^2$. We will now use DDFT
to shed light on these findings.

\subsubsection{Dynamic density-functional theory analysis.}
If we consider that the system evolves adiabatically, the evolution equation of
the local density can be written directly from the equilibrium Helmholtz
free energy functional~\cite{Tarazona1999, Rauscher2010},
\begin{equation}
\gamma \frac{\partial\rho(\vec{r},t)}{\partial
t}=\vec{\nabla}.\left[\rho(\vec{r},t)\vec{\nabla}
\frac{\delta\mathcal{F}[\rho(\vec{r},t)]}{\delta\rho(\vec{r},t)}\right].
\label{TDFT}
\end{equation}
Using the functional defined in Eq. (\ref{funtional}), we obtain a
diffusion equation for the density,
\begin{equation}
\gamma \frac{\partial\rho}{\partial t}=\vec{\nabla}.\left[
A\rho\vec{\nabla}\rho+V_E \kappa cos(\kappa x)\rho\right]+ k_BT\nabla^2 \rho \ \
, \label{evolEq}
\end{equation}
where the first and last terms are related to the particle/particle and
particle/fluid interactions, respectively. They tend to smooth out any spatial
variation of the density. The middle term is related to the particle/field
interaction and promotes the formation of bands. Note that, when one
applies $\vec{\nabla}.$ to the first term one gets  $A\rho \nabla^2 \rho$, a
non-linear diffusion term where the coefficient increases with the density. It
is the interplay of the thermostat temperature, external potential and
particle/particle interactions that controls the kinetics of relaxation and
defines the relaxation time towards the stationary state. A similar diffusion
equation can also be obtained from the Fokker-Planck formalism \cite{Zapperi2001}.

\begin{figure}[t]
\centering
  \includegraphics[width=\columnwidth]{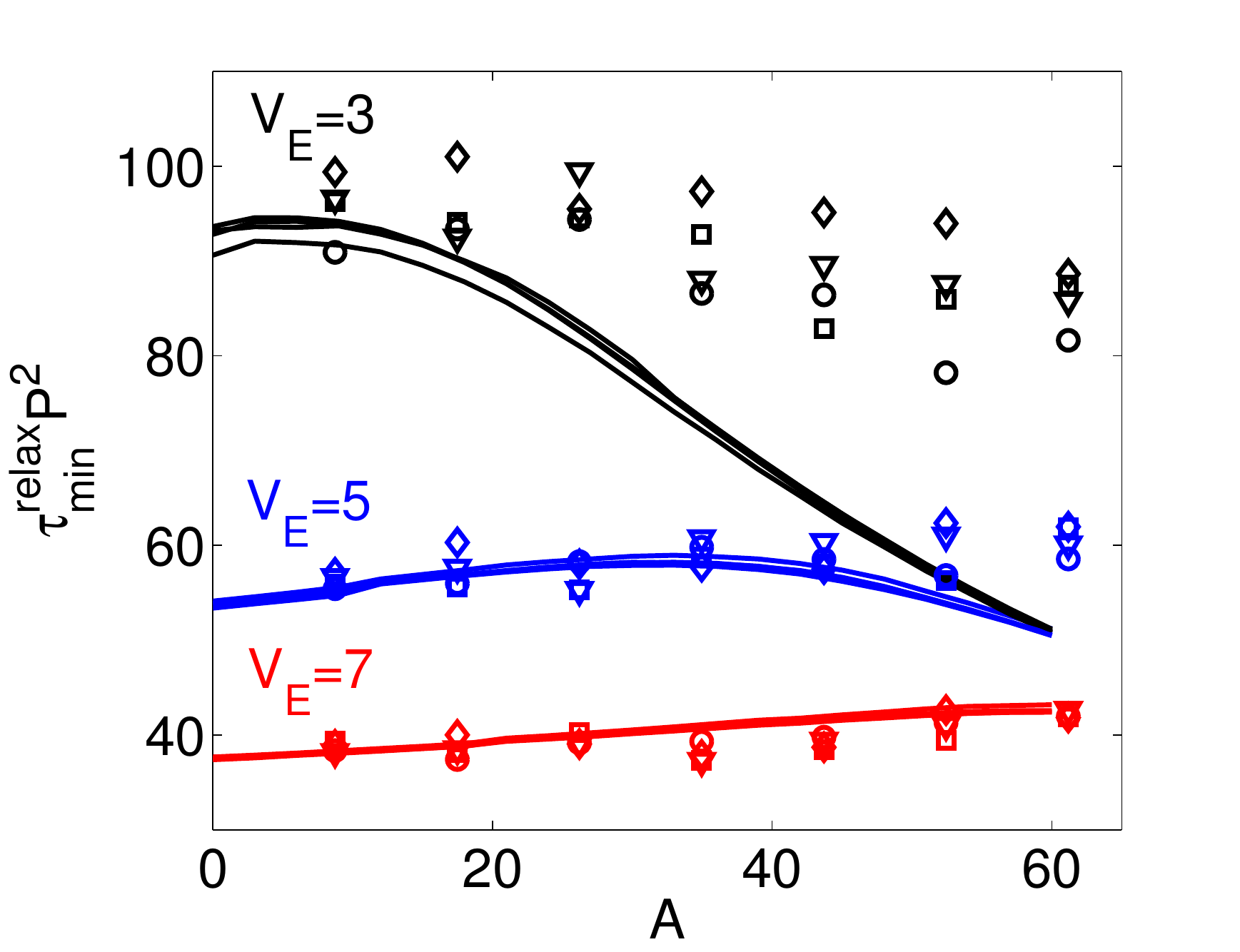}
  \caption{Relaxation time of the minimal density. The scattered data are the
simulation results and the solid lines the ones from the DDFT calculations.
Black, blue and red represent respectively $V_E=\{3,5,7\}$ and the circle,
square, triangle and diamond symbols correspond respectively to $P=\{1,2,3,4\}$. The
simulation results are averages over $45$ samples for $V_E=3$ and $25$ samples
in the other cases. The relaxation time does not vary monotonically with the
interaction strength. It exhibits a maximum that depends on the balance
between $A$ and $V_E$. The relaxation time is proportional to $P^{-2}$.
\label{fgr:taumin}}
\end{figure}

\begin{figure}[t]
\centering
  \includegraphics[width=\columnwidth]{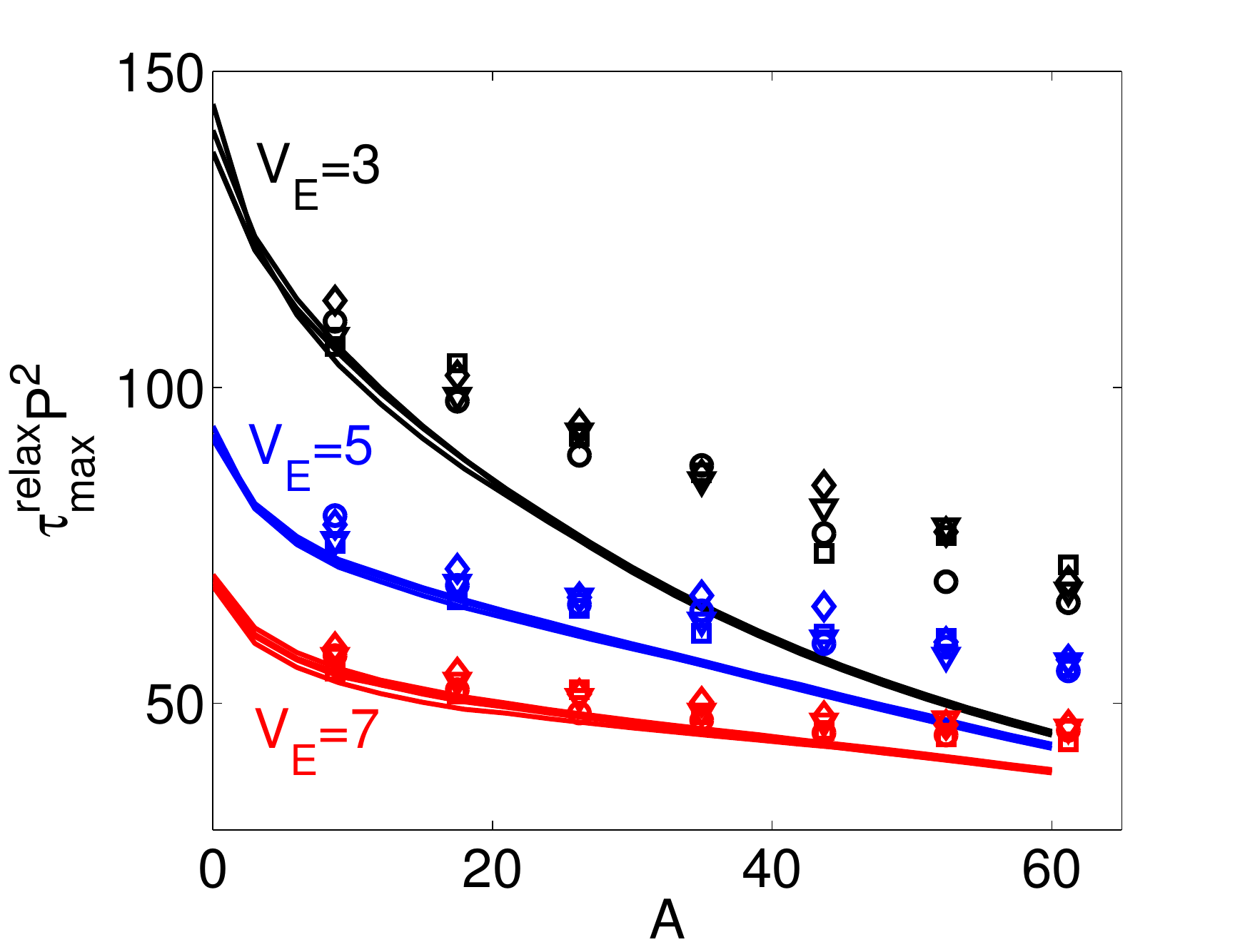}
  \caption{Relaxation time of the maximal density. The scattered data are the
numerical results and the solid lines the ones from the DDFT calculations.
Black, blue and red represent respectively $V_E=\{3,5,7\}$ and the circle,
square, triangle and diamond symbols correspond respectively to $P=\{1,2,3,4\}$. The
simulation results are averages over $45$ samples for $V_E=3$ and $25$ samples
in the other cases. For strong interparticle interactions the relaxation time depends 
weakly on the external potential.
\label{fgr:taumax}}
\end{figure}

We focus on the maximal and minimal densities where $\vec{\nabla} \rho = 0$. Let
us assume that  $\nabla^2 \rho = \rho_2$ is constant in time at these points.
At the minimum $\rho_2 \geq 0$ and at the maximum $\rho_2 \leq 0$. By solving
Eq.~(\ref{evolEq}) we get
\begin{equation}
\rho_{\substack{min \\ max}}(t)=-\frac{k_BT \rho _2}{A\rho _2\mp V_E \kappa
^2}+K_1 \exp \left[ \frac{A \rho _2 \mp V_E \kappa ^2}{\gamma}t \right],
\label{densityMm}
\end{equation}
where the expression with the minus sign is for the density at the minimum and the
plus sign for the density at the maximum. By fixing the initial density it is
possible to calculate the constant
\begin{equation}
K_1 = \frac{N}{L^2}+\frac{k_BT \rho _2}{A\rho _2\mp V_E \kappa ^2}.
\end{equation}
If we define the characteristic relaxation time as
\begin{equation}
\rho(t)-\rho_{\infty}\sim e^{-t/\tau^{relax}}
\end{equation}
where $\rho_{\infty}$ is the density at the given point at equilibrium, we
arrive at the following expression
\begin{equation}
\tau^{relax}_{\substack{min \\ max}}=- \frac{\gamma}{A \rho _2 \mp V_E \kappa
^2}.\label{tau}
\end{equation}
In the region of parameters where $\nabla^2 \rho = const$ the relaxation time is 
inversely proportional to $A$ and $V_E$. The friction coefficient also plays a role 
on the dynamics and the relaxation time is a monotonic increasing function of the viscosity. Replacing
$\kappa$ using Eq.~(\ref{omega}) we conclude that $\tau^{relax} \sim P^{-2}$. The relaxation time decreases 
with the number of bands as, on average, particles have to travel a shorter distance to reach the nearest 
potential minimum. Eq.~(\ref{evolEq}) is solved numerically using first order finite elements in
the 2D square domain with periodic boundary conditions 
\cite{comsol}. Figures~\ref{fgr:taumin}~and~\ref{fgr:taumax} show 
the relaxation time of the minimal and maximal densities, respectively. They collapse for
$P^{-2}$ but the inverse proportionality is only observed for high $A$ which
is where our approximation holds. 

The theoretical results underestimate the relaxation time by comparison with the
data from the Brownian dynamics simulations, consistent with the fact that
DDFT overestimates the diffusion coefficient \cite{Tarazona1999}. The deviation
is greater at low $V_E$ and high $A$, when the particle/particle interactions
dominate the dynamics. Recall that the term for particle/particle interaction in
Eq.~(\ref{funtional}) is the only approximate term.

Note that the relaxation time is always positive. For the minimal density the
curvature $\rho_2$ is typically small and the external potential term dominates.
For the maximal density the curvature is negative and the particle/particle
interaction term dominates over the external potential. In the minimum the
particles try to escape and so the external potential is the relevant
interaction leading to a decrease in the density. However, the process is still
influenced by the presence of neighboring particles and the relaxation 
time increases with $A$ at low $A$, Fig. \ref{fgr:taumin}. For strong interparticle 
repulsions there is a broadening of the
bands; the particles will then travel shorter distances to reach their equilibrium
positions, decreasing $\tau^{relax}$ at the minimal density. By contrast, for
the maximal density the repulsions strongly affect the dynamics. 

The relaxation times obtained from the simulations and the theory coincide for strong potentials as
the effect of the interparticle interactions is irrelevant. 
When $V_E$ decreases the particle/particle repulsion becomes relevant and the LDA approximation
leads to increasing deviations from the simulation results. Equation (\ref{tau})
suggests that for high enough values of $A$ the relaxation times do not depend
on $V_E$ and $P$ and converge to the same value in line with the results shown in Figs. \ref{fgr:taumin} and \ref{fgr:taumax}. For lower $A$ at the maxima, where the particles are compressed, equilibrium takes longer to be reached.

\section{Conclusions~\label{sec::conclusions}}
We studied the dependence of the structure and relaxation dynamics of the
equilibrium state of a colloidal suspension on the external potential and the
interpaticle interactions. The analysis based on density-functional theory was
found to be in good agreement with the results from Brownian dynamics
simulations.

In the stationary state the system exhibits a band-like structure. The
thickness of the bands measured in terms of the mean square displacement,
$\sigma$, decreases with the strength of the potential, which leads in turn to
higher densities around the external potential minima. We also showed that
$\sigma$ scales as $P^{-2}$, where $P$ is the number of bands.  In the limit of
weak external potentials, $\sigma$ increases linearly with the potential
strength, $V_E$.

The relaxation towards the stationary extreme densities is exponential but the
relaxation times are different. The system takes longer to reach equilibrium at
the density maxima except when the external potential becomes significant weak
in which case the relaxation time does not depend on the position. At the
density maxima $\tau^{relax}$ decreases when both $V_E$ and $A$ increase but at
the density minima it displays a more complex behavior and there is a maximum
relaxation time that depends on the balance between $V_E$ and $A$.

Electromagnetic fields are used in the laboratory to promote the formation of a
variety of colloidal structures. Our results quantify the relaxation times and
the equilibrium structures of simple colloidal systems, which exhibit a rich
behaviour that may be controlled in a straightforward manner.

\section{Acknowledgements}
We acknowledge fruitful discussions with D. de las Heras and C. Dias, as well as
financial support from the Portuguese Foundation for Science and Technology
(FCT) under Contracts nos. EXCL/FIS-NAN/0083/2012, UID/FIS/00618/2013, and
IF/00255/2013.

\bibliography{refs}

\begin{thebibliography}{30}
\bibitem [{com()}]{comsol}%
  \BibitemOpen
  \href@noop {} {\enquote {\bibinfo {title} {The \mbox{COMSOL Multiphysics}
  software is used to implement the \mbox{FEM}.}}\ }\BibitemShut {NoStop}%
\end{thebibliography}

\end{document}